\documentclass[12pt]{article}
\usepackage[utf8]{inputenc}
\usepackage{amsmath}
\usepackage{amsfonts}
\usepackage{amssymb}
\usepackage{graphicx}
\usepackage[space]{grffile}
\usepackage{flafter}
\usepackage{multirow}
\usepackage{tikz}
\usetikzlibrary{shapes,arrows}
\usepackage{epsfig,amsfonts,amsthm}
\usepackage[normalem]{ulem}
\usepackage{amsmath,amssymb}
\usepackage{array}
\usepackage{amsmath}
\usepackage{amsfonts}
\usepackage{amssymb}
\usepackage{subfig}
\usepackage{wrapfig}
\usepackage{graphicx}
\usepackage{cite}
\newcommand{\be}{\begin{equation}}
\newcommand{\ee}{\end{equation}}
\newcommand{\bea}{\begin{eqnarray}}
\newcommand{\eea}{\end{eqnarray}}

\usepackage{color}

\def\lsim{\mathrel{\rlap{\lower4pt\hbox{\hskip1pt$\sim$}}
    \raise1pt\hbox{$<$}}}         
\def\gsim{\mathrel{\rlap{\lower4pt\hbox{\hskip1pt$\sim$}}
    \raise1pt\hbox{$>$}}}         

\def\beq{\begin{equation}}
\def\eeq{\end{equation}}
\def\bea{\begin{eqnarray}}
\def\eea{\end{eqnarray}}

\def\<{\left\langle}
\def\>{\right\rangle}

\usepackage[normalem]{ulem}

\def\lsim{\mathrel{\rlap{\lower4pt\hbox{\hskip1pt$\sim$}}
    \raise1pt\hbox{$<$}}}         
\def\gsim{\mathrel{\rlap{\lower4pt\hbox{\hskip1pt$\sim$}}
    \raise1pt\hbox{$>$}}}         

\def\beq{\begin{equation}}
\def\eeq{\end{equation}}
\def\bea{\begin{eqnarray}}
\def\eea{\end{eqnarray}}

\def\<{\left\langle}
\def\>{\right\rangle}

\newcommand{\bt}{\begin{tabular}}
\newcommand{\et}{\end{tabular}}

\usepackage{hyperref}

\hypersetup{
   colorlinks=true,       
   linkcolor=blue,        
   citecolor=red,         
   filecolor=magenta,      
   urlcolor=cyan,           
   linktocpage = true,
   }

\usepackage{tikz}
\usetikzlibrary{decorations.pathmorphing,decorations.markings}

\usepackage{graphicx}

\tikzset{
photon/.style={decorate, decoration={snake,amplitude=2pt, segment length=5pt}, draw=black},
particle/.style={draw=black, postaction={decorate}, decoration={markings,mark=at position .5 with {\arrow[draw=black]{>}}}},
antiparticle/.style={draw=black, postaction={decorate}, decoration={markings,mark=at position .5 with {\arrowreversed[draw=black]{>}}}},
gluon/.style={decorate, draw=black, decoration={coil,amplitude=4pt, segment length=5pt}},
goldstone/.style={draw=green,postaction={decorate},decoration={markings,mark=at position .5 with {\arrow[draw=blue]{>}}}}
}

\allowdisplaybreaks[2]
\begin{document}

\title{\hfill ~\\[-50mm]
                  \textbf{Revisiting Jet Clustering Algorithms \\
for New Higgs Boson
Searches\\ in Hadronic Final States
                }        }
\date{}

\author{\\[-5mm]
A. Chakraborty\footnote{E-mail: {\small\tt amit.c@srmap.edu.in}} $^{1}$,\
S. Dasmahapatra\footnote{E-mail: {\small\tt sd@ecs.soton.ac.uk}} $^{2}$,\
H.A. Day-Hall\footnote{E-mail: {\small\tt hadh1g17@soton.ac.uk}} $^{3,4}$,\
B. Ford\footnote{E-mail: {\small\tt b.ford@soton.ac.uk}} $^{3}$,\ \\
S. Jain\footnote{E-mail: {\small\tt s.jain@soton.ac.uk}} $^{3}$,\
S. Moretti\footnote{E-mail: {\small\tt stefano@phys.soton.ac.uk}}
\footnote{E-mail: {\small\tt stefano.moretti@physics.uu.se}} $^{3,5}$,\
E. Olaiya\footnote{E-mail: {\small\tt emmanuel.olaiya@stfc.ac.uk}} $^{4}$,\
C.H. Shepherd-Themistocleous\footnote{E-mail: {\small\tt claire.shepherd@stfc.ac.uk}} $^{4}$
\\ \\
\emph{\small $^1$Department of Physics, School of Engineering and Sciences, SRM University AP,}\\
\emph{\small  Amaravati, Mangalagiri 522240, India}\\
\emph{\small $^2$School of Electronics and Computer Science, University of Southampton,}\\
\emph{\small Southampton, SO17 1BJ, United Kingdom}\\
\emph{\small $^3$School of Physics and Astronomy, University of Southampton,}\\
\emph{\small Southampton, SO17 1BJ, United Kingdom}\\
\emph{\small  $^4$Particle Physics Department, Rutherford Appleton Laboratory,}\\
\emph{\small Chilton, Didcot, Oxon OX11 0QX, United Kingdom}\\
\emph{\small $^5$ Department of Physics and Astronomy, Uppsala University, Uppsala, Sweden}\\
[3mm]
  }

\maketitle

\vspace*{-10mm}

\begin{abstract}
\noindent
{We assess the performance of different jet-cluster\-ing algorithms, in the presence of different resolution parameters and reconstruction procedures, in resolving fully hadronic final states emerging from the chain decay of the discovered Higgs boson into pairs of new identical Higgs states, the latter in turn decaying into bottom-antibottom quark pairs. We show that, at the Large Hadron Collider (LHC), both the efficiency of selecting the multi-jet final state and the ability to reconstruct from it the masses of the Higgs bosons (potentially) present in an event sample depend stron\-gly on the choice of acceptance cuts,
jet-clustering algorithm as well as its settings. Hence, we indicate the optimal choice of the latter for the purpose of establishing such a benchmark Beyond the SM (BSM) signal. We then repeat the exercise for a heavy Higgs boson  cascading into two SM-like Higgs states, obtaining similar results. }
 \end{abstract}
\thispagestyle{empty}
\vfill
\newpage
\section{Introduction}
The ultimate motivation of our study is to address the incomplete nature of the Standard Model (SM) of particle physics, by indeed  looking for signs of some physics Beyond the SM (BSM). In particular, we pose the question of whether different jet clustering techniques might be more or less suited to find particular final states of interest, specifically, those from topologies involving an extended Higgs sector manifesting itself in cascade decays. In such scenarios, as we will explain in more detail, high $b$-jet multiplicity final states are expected and a point worth addressing is which current experimental jet reconstruction is in fact optimal for these types of searches.

Several BSM scenarios with an enlarged Higgs sector allow for the existence of additional neutral Higgs boson states, CP-even or CP-odd. These resonances can be both lighter, or heavier, than the SM-like Higgs boson  discovered at the LHC in 2012, which has a mass of approximately 125 GeV \cite{Aad:2012tfa}. These new physics frameworks are very common, in particular, in Supersymmetry (SUSY), in both minimal and  non-minimal realisations of it \cite{Book}. Specifically, to find a spectrum where there are lighter or heavier Higgs states than the discovered one, one needs look  no further than the Next-to-Minimal Supersymmetric Standard Model (NMSSM) \cite{Ellwanger:2009dp}. However, if one departs from SUSY and remains with low-energy models, a rather simple BSM framework including these light states in its particle spectrum is  the  2-Higgs Doublet Model (2HDM) \cite{Gunion:1989we,Gunion:1992hs,Branco:2011iw}. 

\begin{figure}[ht!]
	\centering
	\includegraphics[scale=0.7]{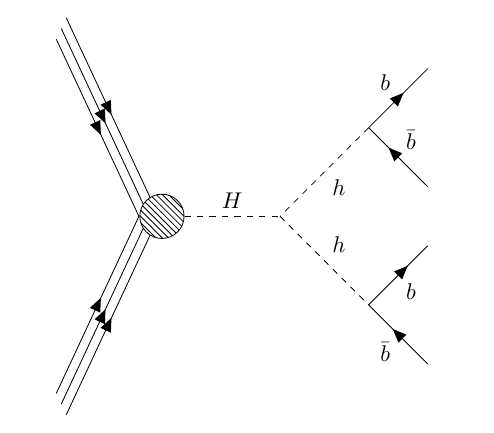}
	\caption{The 2HDM process of interest, where the SM-like Higgs state ($m_H=125$ GeV) produced from gluon-gluon fusion decays into a pair of lighter scalar Higgs states, $hh$, each in turn decaying into $b\bar{b}$ pairs giving a $4b$ final state.}
\label{fig:diagram}
\end{figure}

In such a 2HDM, two complex Higgs doublet fields undergo Electro-Weak Symmetry Breaking (EWSB), yielding five physical Higgs states, labelled as $h$, $H$ (which are CP-even with, conventionally, $m_h<m_H$), $A$ (which is CP-odd) and a pair of charged states $H^\pm$ with mixed CP properties. It is currently possible for the observed 125 GeV Higgs boson to be identified as either $h$ or $H$ in 2HDM scenarios. In both cases, when $m_h<m_H/2$ or $m_A<m_H/2$, the decays $H\to hh$ and/or $H\to AA$ (respectively) may occur. Taking $h$($H$) as the SM-like 125 GeV Higgs boson, for a $H$($h$) state with a mass of order 250(60) GeV or more(less), the dominant decay mode in a 2HDM is bottom-antibottom quark pairs \cite{Moretti:1994ds,Djouadi:1995gv}, i.e., $h\to b\bar b$, so that the final state emerging from the hard scattering $pp\to H\to hh$ is made up, at the partonic level, of four (anti)quarks\footnote{Notice that the same argument can be made for the case of $pp\to H\to AA\to b\bar b b\bar b$ when $m_A<m_H/2$.}, see Fig.~\ref{fig:diagram}. However, due to the confinement properties of Quantum Chromo-Dynamics (QCD), the partonic stage is not accessible by experiment, only the hadronic ``jets" emerging at the end of the parton shower and hadronisation phase are seen.

In order to decipher the source of these showers of ha\-drons, ``jet clustering algorithms" are currently used. Jet clustering algorithms reduce the complexity of such final states by attempting to rewind the showering back to the parton it originated from, such that we move from a large sample of particles to a smaller number, each of which represents a state emerging from the hard interaction of interest in a given event. In other words, we characterise the sample of particles originating from a single parton as an object in itself:  a jet. Needless to say, there is a variety of jet clustering algorithms available and we will dwell at length on these in a forthcoming section. 

The purpose of this paper is to determine whether alternative jet reconstruction tools, in particular a modification to traditional sequential combinations algorithms employing a variable inter-jet  distance measure \cite{Krohn:2009zg} (so-called `variable-$R$' algorithms, where $R$ represents a typical cone size characterising the jet), might be better suited to the $4b$ final states coming from, e.g.,  2HDMs. We approach this problem from a theoretical perspective so as to thoroughly test a range of different combinations, in order to inform whether a detailed experimental analysis might be worth pursuing. Furthermore, the $4b$ final state that we are invoking here is an ubiquitous signal of BSM Higgs boson pairs\footnote{Here, ubiquitous refers to the fact that this signal is very typical of a variety of BSM scenarios, so that we effectively use the 2HDM for illustration purposes. Our results can therefore be applied to the case of other new physics models.}, crucially giving access (through the extraction of the $H$ and/or $h$ state properties) to key features of the underlying BSM scenario, e.g., in the form of the shape of the Higgs potential, hence, of the vacuum stability and perturbative phases of it. 

While the above outlines that the problem of optimal jet reconstruction is clearly an experimental endeavour, we have mentioned that this study is undertaken at a theoretical level. The two aspects are however reconciled in this paper by employing a simplified analysis with respect to the ones carried out on data, yet a rather sophisticated one based on Monte Carlo (MC) event generation,  
in order to reliably compare the relative performance of traditional fixed-$R$ jet clustering against a variable-$R$ method. A comprehensive, more realistic, experimental investigation is left to a future study.  For example, a key feature of the hadronic final state initiated by $b$-quarks that we intend to study  is that the emerging jets can be ``tagged'' as such, unlike the case of lighter (anti)quarks and gluons, which are largely indistinguishable from each other. Here, we implement a simplified method of tagging using MC truth information on the $b$-partons, along with a probabilistic implementation of inefficiencies. For a more detailed discussion on $b$-tagging at detector level, we refer the reader to \cite{Scodellaro:2017wli}. 

{{The layout of the paper is as follows. In the next two sections, we describe how we performed jet reconstruction, $b$-tagging, the tools  used for our simulations  as well as discuss the cutflow adopted.}} In the one following, we present our results for both signal and background. Then, we conclude.

\section{Methodology}
\subsection{Jet Clustering Algorithms}
In order to extract proper physics from hadronic sprays found in particle detectors,  algorithms are used to characterise the detected radiation into distinguishable objects, the aforementioned jets.
There is indeed a rich history associated with the development and evolution of algorithms for jet definition, beginning in 1977 with Sterman and Weinberg \cite{Sterman:1977wj}, initially deployed in the context of $e^+ e^- \rightarrow$ hadron scatterings. For a detailed look into the evolution of jet clustering, we refer the reader to \cite{Moretti:1998qx}. Here we focus on the modern implementations.

The type of algorithms currently utilised are known as sequential recombination algorithms, or jet clustering algorithms  \cite{Moretti:1998qx}. Rather than categorising the entire event at once, as in previous approaches, each particle in the event begins {with a separate, intermediate grouping, a pseudojet. These initial pseudojets are iteratively combined to form larger pseudojets} based on some inter-particle distance measure, until all particles are gathered into stable jets.

The type of algorithms currently deployed at the LHC all take a similar form, descending from the generalised $k_T$ algorithm used initially in $e^+e^-$ colliders. This uses an inter-particle distance measure which can be written in the form
\begin{equation}\label{eqn:dij}
d_{ij} = \min(p_{Ti}^{2n},p_{Tj}^{2n})\Delta R^2_{ij},
\end{equation}
where $\Delta R^2_{ij} =(y_i - y_j)^2 + (\phi_i - \phi_j)^2$ is the angular separation between {pseudojets $i$ and $j$}, with $y$ and $\phi$ being the rapidity and azimuthal angle of the associated final state hadron. They also make use of the `beam distance', which is the separation between {pseudojet} $i$ and the beam $B$,
\begin{equation}
{d_{Bi} = p^{2n}_{Ti}R^2}.
\end{equation}
Note that we use the same notation as in \cite{Krohn:2009zg}, where $R^2$ is included in the definition of $d_{Bi}$. (An alternative convention is to embed $R^2$ into the definition of $d_{ij}$ such that $d_{ij} = \min(p_{Ti}^{2n},p_{Tj}^{2n})\frac{\Delta R^2_{ij}}{R^2}$, leaving $d_{Bi} = p^{2n}_{Ti}$, like in \cite{Cacciari:2011ma}.) For a set of {pseudojets}, all possible $d_{ij}$'s and $d_{Bi}$'s are calculated and the minimum is taken. If the minimum is a $d_{ij}$, {pseudojets} $i$ and $j$ are combined and the process is repeated. If, instead, a $d_{Bi}$ is the minimum, then $i$ is declared a jet and removed form the sample. This procedure is then  repeated until all objects are classified into jets.

In $d_{Bi}$, $R$ is a fixed input variable which dictates the size of the jet and acts as the cut-off for any particle pairing. If we consider some pair of {pseudojets} $i$ and $j$, with $i$ having lower $p_T$ (and hence being selected in $d_{ij}$), we can write (for $n\ge0$)
\begin{equation}
{d_{ij} = \Delta R^2_{ij} p^{2n}_{Ti} = \frac{\Delta R^2_{ij}}{R^2} d_{Bi}.}
\end{equation}
Since we require the ratio $\frac{\Delta R^2_{ij}}{R^2}<1$ to avoid declaring $i$ a jet over merging $i$ with $j$, we can see that $R$ acts as an effective cut off for the maximum separation of two pseudojets to be combined and, hence, it is proportional to the final jet size.
From this general formulation, the main two jet clustering algorithms  currently in use at the LHC are Cambridge-Aachen (CA) \cite{Wobisch:1998wt,Dokshitzer:1997in} ($n=0$) and anti-$k_T$\cite{Cacciari:2008gp,Salam:2009jx} ($n=-1$).

\subsection{Jet Clustering with Variable-$R$} \label{JetClusteringwithVariableR}
Both the anti-$k_T$ and CA algorithms require an input parameter $R$ for the anti-$k_T$ algorithm, this can be interpreted as the radius of the cone formed by tracing the jet back to the interaction point\footnote{It is an important distinction to notice that $R$ only looks like a cone parameter.
    {
    In any generalised $k_T$ algorithm, the migration of the pseudojet early in the clustering
    allows the jets with a radius of up to $1.5R$ to form.
    However, as the pseudojet grows, new additions no longer cause significant movement.
For the anti-$k_T$, the `min' in Eq. (\ref{eqn:dij}) picks out the larger $p_T$ pseudojet.
If we take $i$ as the higher $p_T$ object, then, if $j$ is further from $i$ than $R$, they will not combine.
In other words, if $i$ is close to the eventual jet axis, and no longer undergoes significant movement,
$R$ effectively ends up as the jet radius.
While this point is important, referring to $R$ as an effective cone size is largely justified.}}.
Recall this acts as a cut off for combining hadrons and can therefore be interpreted as implementing a size limit on the jets depending on the particle separation.

The angular spread of the final jet constituents has a dependence on the $p_T$ of
the initial partons. For higher $p_T$ objects the decay products
will be packed into a more tightly collimated cone whereas for
low $p_T$ objects one would expect the resulting jet constituents to
be spread over some wider angle. One can therefore
imagine the need for carefully selecting the $R$ value used for
clustering depending on the $p_T$ of the final state jets, but what
about in a multi-jet scenario where the final state partons have a
wide range of $p_T$'s?

The aforementioned variable-$R$ jet clustering algorithm \cite{Krohn:2009zg} alters the above scheme so as to adapt to events with jets of varying cone size. A modification to the distance measure $d_{ij}$ is made, by replacing the fixed input parameter $R$ with a $p_T$ dependent $R_{\text{eff}}(p_T) = \frac{\rho}{p_T}$, where $\rho$ is a chosen dimensionful constant (taken to be ${\cal O} (\text{jet}\ p_T)$). With this replacement, the beam distance measure becomes
\begin{equation}
{d_{Bi} = p^{2n}_{Ti}R_{\text{eff}}(p_{Ti})^2.}
\end{equation}
When the distance measures are calculated, $d_{Bi}$ will therefore be suppressed for objects with larger $p_T$ and hence these objects become more likely to be classified as jets. For low $p_T$ objects, $d_{Bi}$ is enhanced and so these are more likely to be combined with a near neighbour, thus increasing the spread of constituents in the eventual jet.

In this approach, the clustering process is modified, so that one can more efficiently avoid events with very wide jets at low $p_T$. The parameter $\rho$ can be optimised to obtain the maximum desired sensitivity. Similarly, this can be done for other parameters too, such as $R_{max/min}$, which are cut offs for the maximun and minimum allowed $R_{\text{eff}}$, i.e., if a jet has $R_{\text{eff}} > R_{max}$, it is overwritten and set to $R_{\text{eff}} = R_{max}$ and equivalently for $R_{min}$.

In multijet signal events where one might expect signal $b$-jets with a wide spread of different $p_T$'s, we hypothesise that a variable-$R$ reconstruction procedure could improve upon the performance of traditional fixed-$R$ routines. In particular, using a variable-$R$ alleviates the balancing act of finding a single fixed cone size that suitably engulfs all of the radiation inside a jet, without sweeping up too much outside `junk'. 

\begin{figure}[h!]
	\begin{center}
	\includegraphics[scale=0.42]{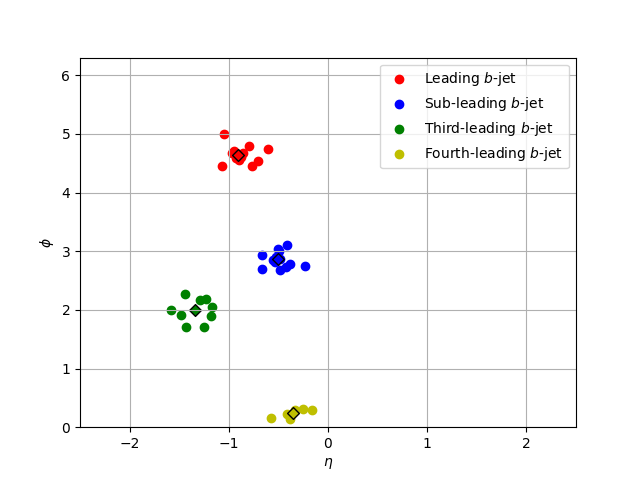}
	\includegraphics[scale=0.42]{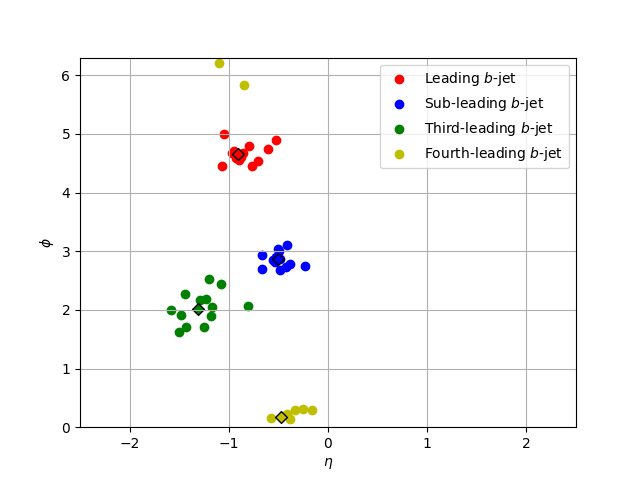}
	 \end{center}
	\caption{The same MC event in $(\eta , \phi)$ space is shown in both plots. Tracks have been clustered with (upper) a fixed $R=0.4$ and (lower) the variable-$R$ algorithm. The coloured points are the constituents of the corresponding $b$-jet in the legend and black outlined diamonds are at the overall $(\eta , \phi)$ coordinates of the $b$-jet $4$-momentum.}
\label{fig:event_images}
\end{figure}

\begin{figure}[h!]
	\begin{center}
	\includegraphics[scale=0.42]{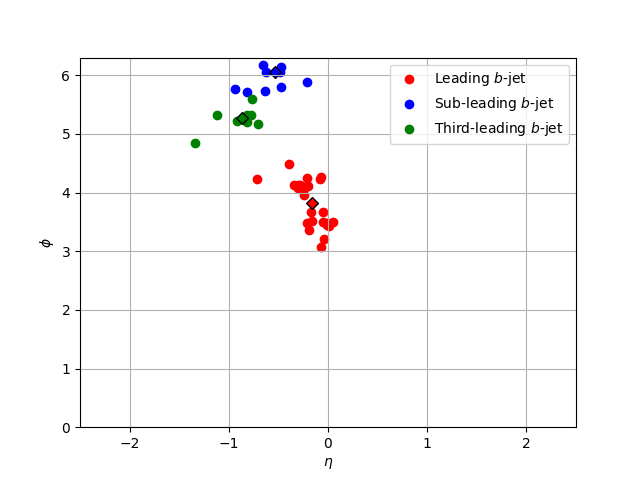}
	\includegraphics[scale=0.42]{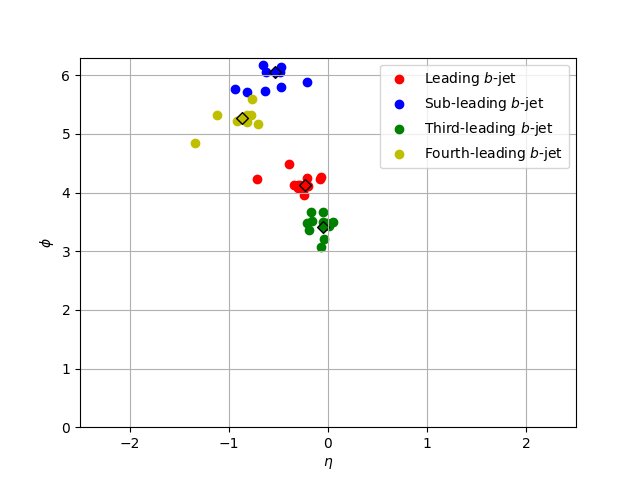}
	 \end{center}
	\caption{As in Fig.~\ref{fig:event_images}, a different event is clustered with a fixed $R=0.8$ (upper) and the variable $R$ algorithm (bottom).  This results in the formation of three $b$-jets for fixed $R$ (upper) and four $b$-jets for the variable-$R$ algorithm (lower). }
\label{fig:event_images2}
\end{figure}

To illustrate this, we show in Fig.~\ref{fig:event_images} the constituents of $b$-tagged jets (hereafter, $b$-jets for short) in the same event, which have been clustered using both a
variable-$R$ and fixed $R=0.4$ scheme.  While the contents of the leading and sub-leading $b$-jets are comparable, the variable-$R$ jets gather a wider cone of constituents for lower $p_T$ jets. The loss of constituents restricts our ability to accurately reconstruct Higgs masses when analysing $b$-jets. In
Fig.~\ref{fig:event_images2} we see an event where only three $b$-jets are resolved by a larger fixed cone ($R=0.8$) algorithm.  In contrast, all four of the expected jets forming the signal are reconstructed by variable-$R$. While fixed-$R$ sweeps radiation from a nearby jet into the leading $b$-jet, variable-$R$ is able to resolve both due to the the smaller effective radius $R_{\text{eff}}$ induced by the larger $p_T$ of the leading $b$-jet.  At the same time  $R_{\text{eff}}$ adapts to a larger value to suitably reconstruct the lower
$p_T$ jets.

\subsection{Implementation of $b$-Tagging}\label{btagging}
In this paper we implement a simplified MC informed $b$-tagger. For events clustered using a fixed-$R$ cone size, jets within angular distance $R$ from a parton level $b$-(anti)quark are searched for and tagged as appropriate. For scenarios where multiple jets are found, the closest is taken as the assignee for the $b$-tag. When the variable-$R$ approach is used, the size of the tagging cone is taken as the effective size $R_{\text{eff}}$ of the jet.

In addition, we account for the finite efficiency of identifying a $b$-jet as well as the non-zero probability that $c$-jets and light-flavour plus gluon jets are mis-tagged as $b$-jets. We apply the variable mis-tag rates and tagging efficiencies from a specific detector card (see below).

\subsection{Simulation Details}
We consider two sample Benchmark Points (BPs) that we call BP1 and BP2.  In BP1, we take the lighter Higgs to be the SM-like Higgs Boson with $m_h=125$ GeV and we set $m_H=700$ GeV.  In BP2, the heavier Higgs has $m_H=125$ GeV with $m_{h}$ set to 60 GeV. Both BPs are in a 2HDM Type-II (2HDM-II henceforth) and have been tested (and passed as not currently excluded) against theoretical and experimental constraints by using 2HDMC \cite{Eriksson:2010zzb}, HiggsBounds \cite{Bechtle:2013wla}, HiggsSignals \cite{Bechtle:2013xfa} as well as checking flavour constraints with SuperISO \cite{Mahmoudi:2009zz}. We generate samples of $\mathcal{O}(10^5)$ events, with $\sqrt{s}$ = 13 TeV for the LHC energy.  Concerning the latter code, we have tested our BPs  against the following flavour constraints on meson decay Branching Ratios (BRs) and mixings, all to the $2\sigma$ level: ${\rm BR}(b \rightarrow s \gamma)$, ${\rm BR}(B_s \rightarrow \mu \mu)$, ${\rm BR}(D_s \rightarrow \tau \nu)$, ${\rm BR}(D_s \rightarrow \mu \nu)$, ${\rm BR}(B_u \rightarrow \tau \nu)$, $\frac{{\rm BR}(K \rightarrow \mu\nu)}{{\rm BR}(\pi \rightarrow \mu \nu)}$, ${\rm BR}(B \rightarrow D_0 \tau \nu)$ and $\Delta_0(B \rightarrow K^* \gamma)$.

The production and decay rates for the subprocesses $gg$, $q\bar q$ $\to H\to hh\to b\bar b b\bar b$
are presented in Tab.~\ref{tab:params}, alongside the 2HDM-II input parameters. (Notice that the $H$ and $h$ decay widths are of order MeV: since this is much smaller than the detector resolutions in two-jet and four-jet invariant masses, the Higgs states can essentially be treated as on-shell.) In the calculation of  the overall cross section, the  renormalisation and factorisation scales were both set to be $H_{T}/2$, where $H_T$ is the sum of the transverse energy of each parton. The Parton Distribution Function (PDF) set used was ${\tt NNPDF23_{\bf -}lo_{\bf -}as_{\bf -}0130_{\bf -}qed}$ \cite{Ball:2014uwa}. Finally, in order to carry out a realistic MC simulation, the toolbox described in Fig.~\ref{fig:toolbox} was used to generate and analyse events \cite{Alwall:2014hca,Sjostrand:2007gs,Conte:2012fm,Conte:2018vmg}\footnote{Note that we use the Leading Order (LO) normalisation for the signal cross sections here, to be consistent with the LO implementation of the background cross sections below. This affects our  final results on event rates and significances. However, the main purpose of our paper is to assess the performances  and consequences of jet clustering algorithms, which should be unaffected by the exact values of signal and backgrounds rates.}.

\begin{table}[!h]
\begin{center}

\scalebox{0.6}{
\begin{tabular}{ |c|c|c|c|c|c|c|c|c|c| }
 \hline
 Label & $m_h$ (GeV) & $m_H$ (GeV) & $\tan\beta$ & $\sin (\beta -\alpha)$ & $m_{12}^2$ & BR($H\rightarrow hh$) & BR($h\rightarrow b\overline{b}$)&$\sigma$(pb)  \\
 \hline
BP1 & 125 & 700.668 & 2.355& -0.999 & 1.46$\times 10^5$ &  6.218$\times 10^{-1}$ &6.164$\times 10^{-1}$&1.870$\times 10^{-2}$\\
 \hline
BP2 & 60 & 125 & 1.6 & 0.1 & 4$\times 10^3$ & 6.764$\times 10^{-1}$ &8.610$\times 10^{-1}$&6.688\\
\hline
\end{tabular}
}
\caption{\label{tab:params} The 2HDM-II  parameters and cross sections of the process in Fig. \ref{fig:diagram} for each BP.}
\end{center}
\end{table}

\vspace*{1em}
\tikzstyle{node} = [rectangle, rounded corners, minimum width=3cm, minimum height=1cm,text centered, draw=black]
\tikzstyle{arrow} = [thick,->,>=stealth]

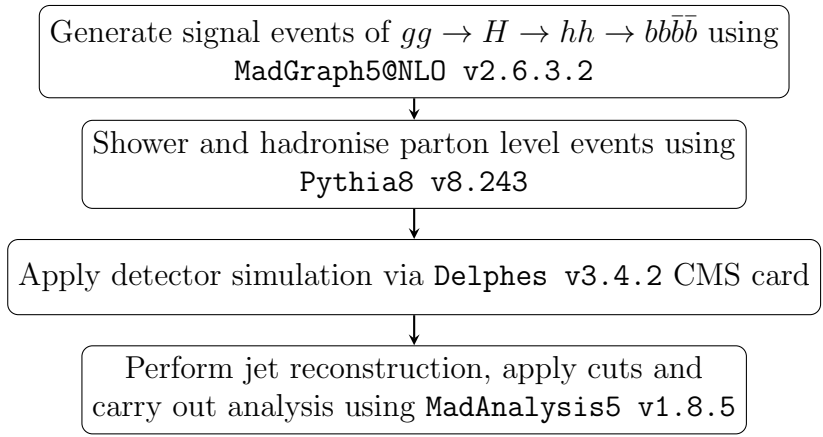
\begin{figure}[htb!]
\centering
\hspace*{-0.5truecm}
\begin{tikzpicture}[node distance=1.5cm]
\node (step1) [node, align=center] {Generate signal events of $gg \rightarrow H \rightarrow h h \rightarrow b b \bar{b} \bar{b}$ using \\ {\tt MadGraph5@NLO v2.6.3.2}};
\node (step2) [node, align=center, below of=step1] {Shower and hadronise parton level events using\\ {\tt Pythia8 v8.243}};
\node (step3) [node, align=center, below of=step2] {Apply detector simulation via {\tt Delphes v3.4.2} CMS card};
\node (step4) [node, align=center, below of=step3] {Perform jet reconstruction, apply cuts and \\ carry out analysis using {\tt MadAnalysis5 v1.8.5}};

\draw [arrow] (step1) -- (step2);
\draw [arrow] (step2) -- (step3);
\draw [arrow] (step3) -- (step4);
\end{tikzpicture}
\caption{Description of the procedure used to generate and analyse MC events. }
\label{fig:toolbox}
\end{figure}

\section{Cutflow}
The introduction of the full  sequence of cuts that we have adopted here requires some justification. 
In existing four $b$-jet analyses by the ATLAS and CMS collaborations that seek to
extract chain decays of Higgs bosons from the background, restrictive cuts have been used for the
ensuing fully hadronic signature.  For BP1, $p_T$ cuts on $b$-jets informed by choices made at CMS~\cite{Sirunyan:2018tki} of all four $b$-jets satisfying $p_T>$50 GeV are used (at trigger level).  Upon enforcing the same cuts as in CMS~\cite{Sirunyan:2018tki} on BP2 for Run 2 and 3 luminosities, we found the signal selection
efficiency  to be too low to create a MC sample suitable for phenomenological analysis. Instead, we have used here a flat cut on all four $b$-tagged jets of 20 GeV, whose viability at the LHC remains to be seen. We indeed provide results in this regime to demonstrate the utility of using a variable-$R$ jet reconstruction algorithm on low-$p_T$ jets from 2HDM-II decays into $bb\bar{b}\bar{b}$ final states. 

\vspace*{1em}
\tikzstyle{node} = [rectangle, rounded corners, minimum width=3cm, minimum height=1cm,text centered, draw=black]
\tikzstyle{arrow} = [thick,->,>=stealth]

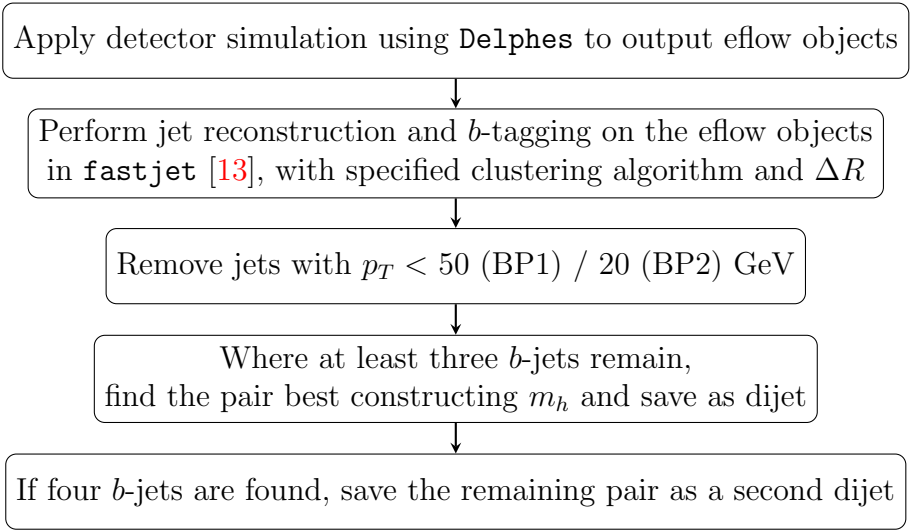
\begin{figure}[!t]
\centering

\begin{tikzpicture}[node distance=1.5cm]
\node (cut1) [node, align=center] {Apply detector simulation using {\tt Delphes} to output eflow objects};
\node (cut2) [node, align=center, below of=cut1] {Perform jet reconstruction and $b$-tagging on the eflow objects\\ in {\tt fastjet} \cite{Cacciari:2011ma}, with specified clustering algorithm and $\Delta R$};
\node (cut3) [node, align=center, below of=cut2] {Remove jets with $p_T <$ 50 (BP1) / 20 (BP2) GeV};
\node (cut4) [node, align=center, below of=cut3] {Where at least three $b$-jets remain, \\ find the pair best constructing $m_h$ and save as dijet};
\node (cut5) [node, align=center, below of=cut4] {If four $b$-jets are found, save the remaining pair as a second dijet};

\draw [arrow] (cut1) -- (cut2);
\draw [arrow] (cut2) -- (cut3);
\draw [arrow] (cut3) -- (cut4);
\draw [arrow] (cut4) -- (cut5);
\end{tikzpicture}
\caption{Description of our initial procedure for jet clustering, $b$-tagging and selection of jets. 
Note that the bulk of our analysis is performed at particle rather than detector level, so MC truth  information is used for cuts on jet constituents.}
\label{fig:cutflow}
\end{figure}
\section{Results}
In this section we present the results for our signal at both the parton and detector level. In the latter case, we also discuss the dominant backgrounds, due to QCD $4b$ production, $gg,q\bar q\to Zb\bar b$ and $gg,q\bar q\to t\bar t$\footnote{In fact, we have checked that the additional noise due to
$t\bar t b\bar b$ events as well as hadronic final states emerging from $W^+W^-, W^\pm Z$ and $ZZ$ production and decay are negligible, once mass reconstruction around $m_h$ and $m_H$ is enforced.}.

\subsection{Parton Level Analysis}

\begin{figure}[h!]
	\begin{center}
	\includegraphics[scale=0.4]{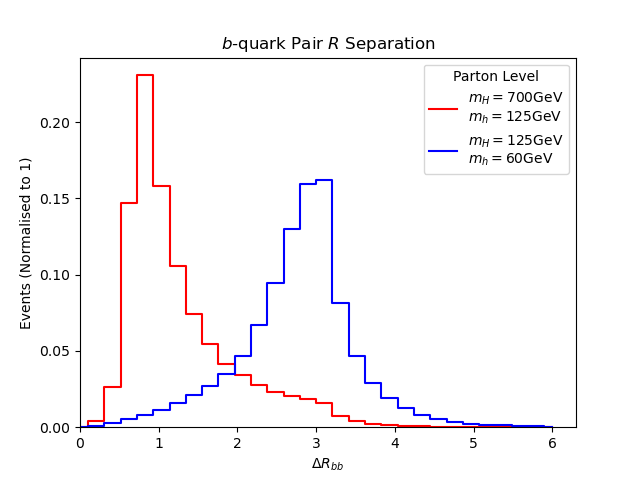}\\
	\includegraphics[scale=0.4]{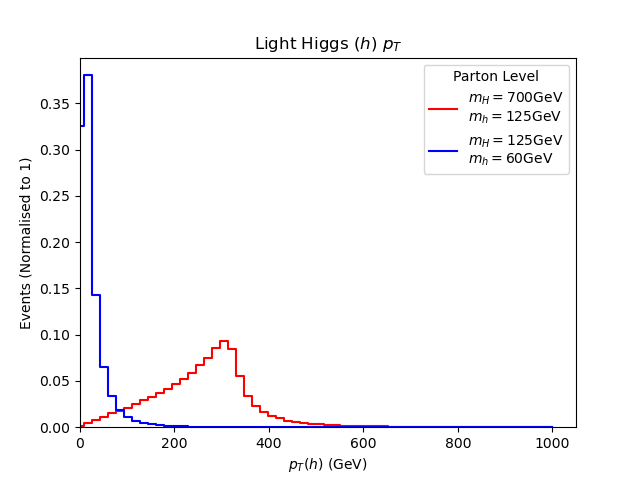}
	\includegraphics[scale=0.4]{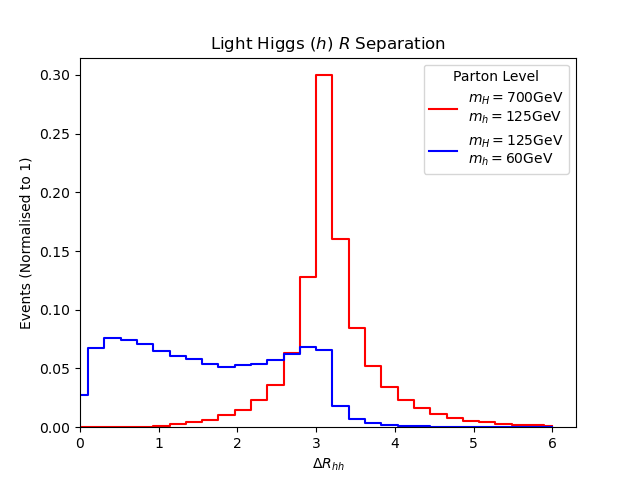}
	 \end{center}
	\caption{Upper panel: the $\Delta R$ distribution between the
two $b$-partons originating from the same $h$. Middle panel:
the $p_T$ distribution of the light Higgs boson $h$ originating from $H$ decay.
Lower panel: the $\Delta R$ distribution between the two $h$ states originating from the
$H$ decay. No (parton level) cuts have been enforced here.}
\label{fig:parton_higgs}
\end{figure}

At the Matrix Element (ME) level, all the events have four $b$-quarks originating
from the decay of the two light Higgs bosons ($h$). We plot the $R$ separation between the $b$-quarks coming from
the same light Higgs state (see upper panel of Fig.~\ref{fig:parton_higgs}).  The two distributions corresponding to BP1 and BP2 are markedly different. 
In general, the angular separation between the decay products $a$ and $b$  in the resonant process
 $X \to a b$ can be approximated as $\Delta R (a,b) \sim \frac{2 m_{X}}{p^{X}_T}$. Hence, we plot in the middle panel of Fig.~\ref{fig:parton_higgs} the transverse momentum of each of the $h$ bosons.
For $m_h$ = 60 GeV, the light Higgs boson has less $p_T$ than for lower values of $m_h$ (owing to the smaller $m_H-m_h$ mass difference). Therefore, the $b$-quarks are more widely separated in this case, compared to $m_H=700$ GeV. In the light of this, we can already conclude that there is a strong correlation between the mass difference $m_H - m_h$ and the cone size of the jet clustering algorithm that ought to be used.  For different choices of the mass of the light Higgs boson, to maximise the number of jets\footnote{This is done also with a view to background rejection.} formed by a clustering algorithm it may be necessary to vary the jet radius parameter, instead of having a fixed radius.   In the lower panel of Fig.~\ref{fig:parton_higgs}, we plot the $\Delta R$ separation between the two light Higgs states. For the configuration in BP1, it is clear (since $\Delta R\approx \pi$) that the $H \to h h $ decay is predominantly back-to-back (in the laboratory frame). However, for $m_h = 60$ GeV,  there is a double peak structure. This occurs due to a recoil effect from Initial State Radiation (ISR), which only becomes apparent at the mass boundary where $m_H \simeq 2m_h$. The inability of the two emerging $h$ states to fly apart implies some overlapping of momenta from $b$-quark showers. Hence, we expect that the output of the clustering algorithm will have a high $b$-jet multiplicity as long that the two $b$-jets stemming from $h$ decays are resolved.  This can depend on detector acceptance and signal selection cuts.
Since $m_h$ is small compared to typical jet $p_T$ thresholds used in applying $b$-tagging, the multiplicity of jets can be reduced.  We will investigate this later.

\begin{figure}[h!]
	\begin{center}
	\includegraphics[scale=0.4]{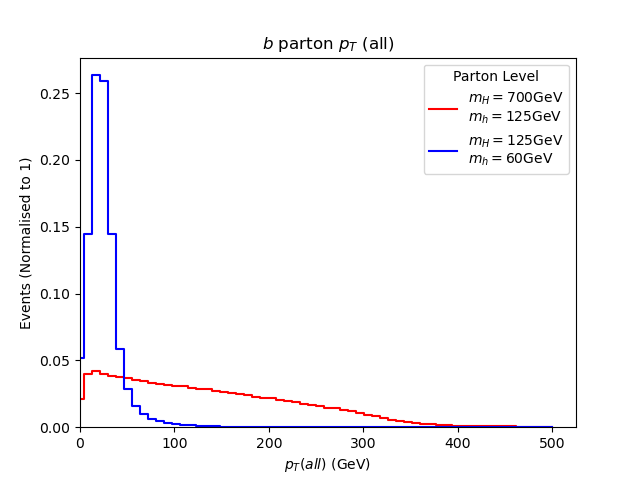}\\
	\includegraphics[scale=0.4]{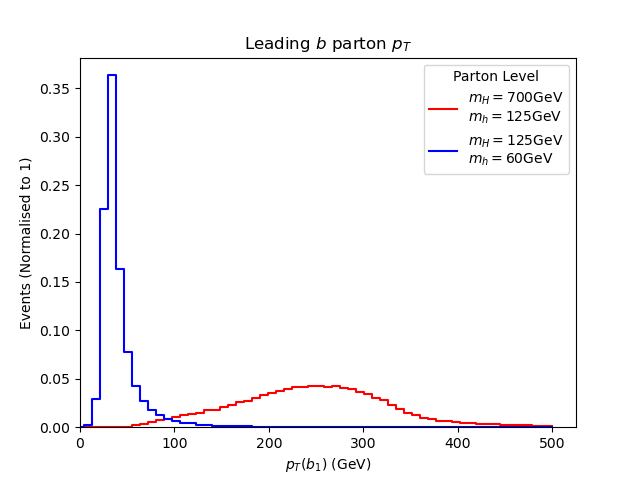}
	\includegraphics[scale=0.4]{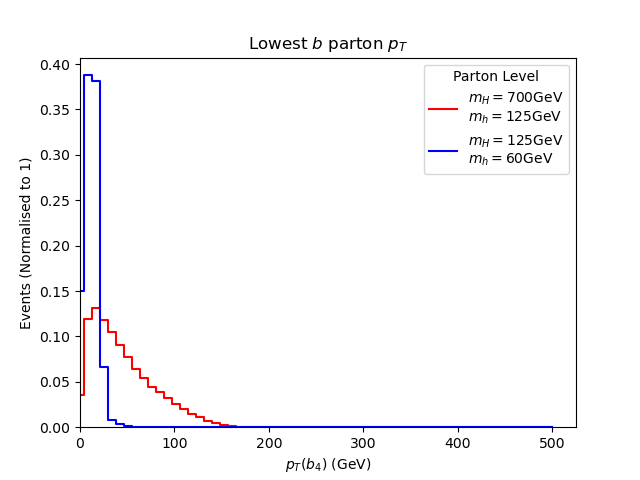}
	 \end{center}
	\caption{Upper panel: the $p_T$ distribution for all $b$-quarks. Middle panel: highest $p_T$ amongst the $b$-quarks. Lower panel: lowest $p_T$
amongst the $b$-quarks. No (parton level) cuts have been enforced here.}
\label{fig:parton_b}
\end{figure}

For a final motivating study to guide the clustering of detected particles, we plot $b$-quark $p_T$ distributions in Fig.~\ref{fig:parton_b}.  From the top histogram find that the $p_T$'s of $b$-quarks span the range of possible values for both mass configurations.  Hence, we would expect the resulting jets to have a similar kinematic spread. In particular, we also plot the highest and lowest $p_T$'s amongst the $b$-quarks in a given event (middle and lower frames, respectively): notice a stark difference in both cases. Further to the discussion in Sec.~\ref{JetClusteringwithVariableR}, one would therefore expect the resulting spread of radiation from each signal $b$-quark to vary in  solid angle and hence the resulting jets to be of differing sizes. This thus motivates the need for a jet reconstruction sequence that adapts to jets of various cone sizes. Therefore, in the next section, we firstly test how jet clustering with fixed-$R$ input behaves and then introduce the variable-$R$ algorithm performance.

\subsection{Jet Level Analysis}
In this section we consider a jet level analysis, using hadronised parton showers that have been run through detector simulation and clustered into jets. We will compare the kinematic distributions of final state $b$-jets, when clustered with a fixed cone as well as with  variable-$R$, for both mass configurations in BP1 and BP2. In particular, we will be interested in the $b$-jet multiplicity, that is the number of $b$-tagged jets in a given event. This is of course indicative of how well our clustering is performing, in that we know the final state has four $b$-quarks, hence, a good algorithm should recover all four. We will also investigate the mass distributions of $b$ dijets and four $b$-jet masses, which indicate our ability to observe the signals containing BSM Higgs Bosons.

We first consider the effect of a variable versus a fixed cone algorithm by observing kinematic variables from signal events for each BP. We choose a value of $R=0.4$, and use the anti-$k_T$ algorithm throughout. (The results for the CA scheme are very similar, so we refrain from presenting these.) For variable-$R$, we use  $\rho=100$ GeV for BP1  and $\rho=20$ GeV for BP2. These values are informed by the $p_T$ scale of the fixed cone $b$-jets. Finally, we use $R_{\text{min}}=0.4$ and $R_{\text{max}}=2.0$ throughout wherever variable-$R$ is used.

\begin{figure}[h!]
	\begin{center}
	\includegraphics[scale=0.4]{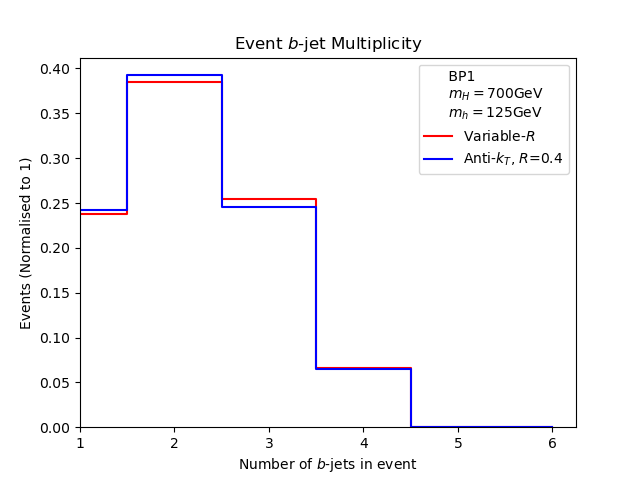}
	\includegraphics[scale=0.4]{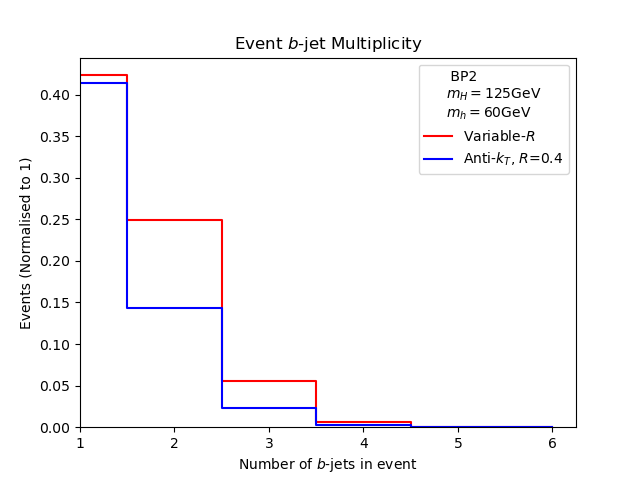}
	 \end{center}
	\caption{Upper panel: The $b$-jet multiplicity distributions for BP1. Lower panel: For BP2.}
\label{fig:bjet_mult}
\end{figure}

We show in Fig.~\ref{fig:bjet_mult} the $b$-jet multiplicity for each of the benchmarks/algorithms.
The stark difference between the two plots in Fig.~\ref{fig:bjet_mult} is due to the relative kinematics of the final state $b$-jets. Due to the different mass configurations, $b$-jets from BP2 have significantly lower $p_T$ than those from BP1, so that significantly more are lost to the trigger as well as from the ($p_T$ dependent) $b$-tagging efficiencies. The use of the variable-$R$ algorithm results in a small increase in events reconstructed with higher $b$-jet multiplicity for BP1 while a more significant shift is evident for BP2.

In order to extract evidence of new physics from $b$-jet signals, we look at the invariant mass of pairs of clustered jets (dijets), in order to reconstruct the mass of the resonance from which they originated.
\begin{figure}[h!]
	\begin{center}
	\includegraphics[scale=0.4]{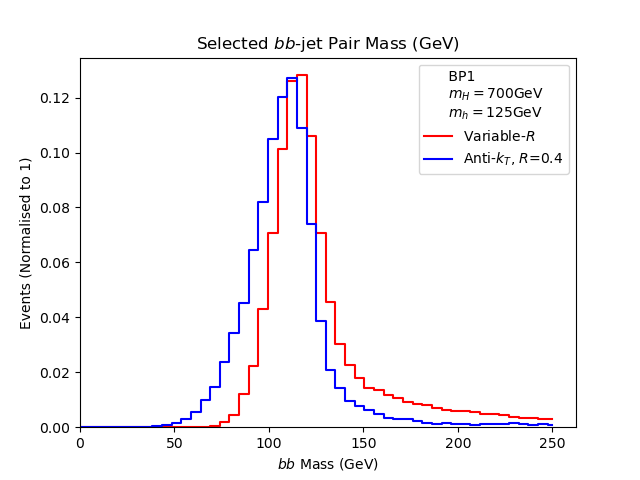}
	\includegraphics[scale=0.4]{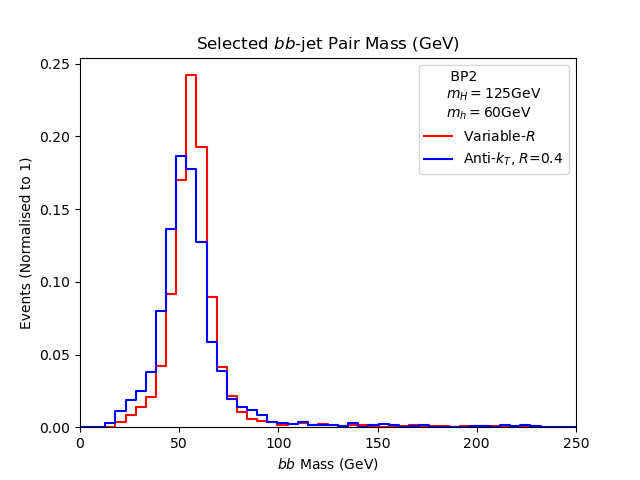}
	 \end{center}
	\caption{The four $b$-jet invariant mass $m_h$ distributions from dijets for BP1 (upper panel) and BP2 (lower panel). The peak of the mass distribution obtained from the variable $R$ algorithm is closer to the MC truth value of the corresponding Higgs.}
\label{fig:bb_mass}
\end{figure}

From Fig.~\ref{fig:bb_mass}, we can see more definitively the benefits of using a variable-$R$ jet clustering algorithm. The four $b$-jet invariant mass $m_H$ distributions from jet clustering for BP1 (upper panel) and BP2 (lower panel) are shown. The peak of the mass distribution obtained from the variable $R$ algorithm is closer to the MC truth value of the corresponding Higgs resonance. 

\begin{figure}[h!]
	\begin{center}
	\includegraphics[scale=0.4]{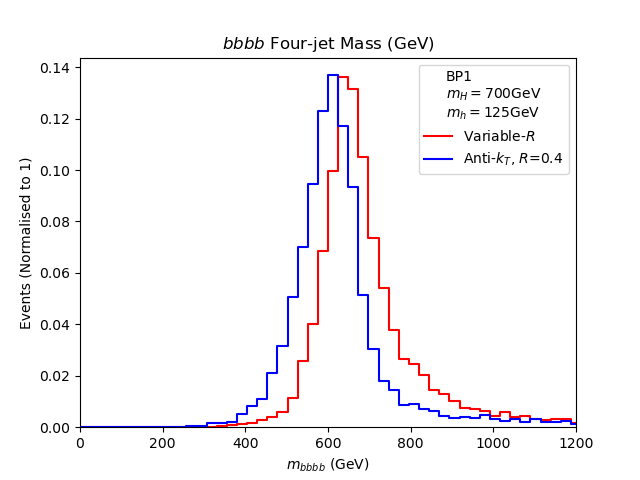}
	\includegraphics[scale=0.4]{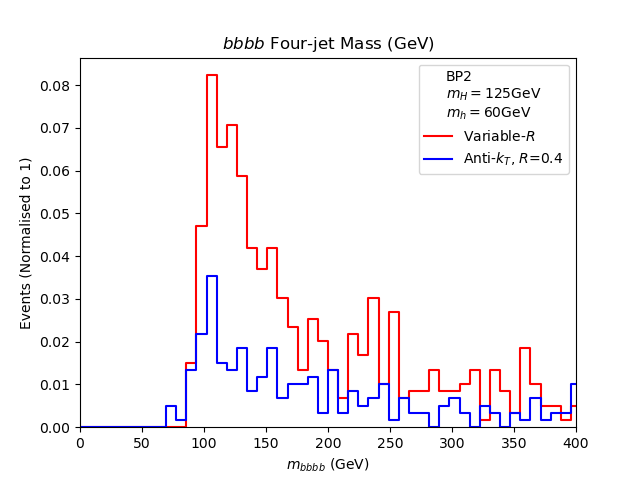}
	 \end{center}
	\caption{The four $b$-jet invariant mass $m_H$ distributions from four jets obtained from jet clustering for BP1 (upper panel) and BP2 (lower panel).}
\label{fig:4b_mass}
\end{figure}

The same behaviour for four $b$-jets masses (Fig. \ref{fig:4b_mass}) can be seen as in the dijet plot, events clustered with variable-$R$ have the $4b$ invariant mass more closely aligned with the expected positions at $m_H$. 

\subsection{Signal-to-Background Analysis}
\begin{figure}[!h]
\centering

\begin{tikzpicture}[node distance=1.5cm]
\node (cut1) [node,align=center] {Select events that contain exactly four $b$-jets};
\node (cut2) [node, align=center,below of=cut1] {Remove event if $|m_{bbbb} - m_H|>$ 50 GeV};
\node (cut3) [node,align=center, below of=cut2] {Using di-jet pairings chosen in above analysis};
\node (cut4) [node, align=center,below of=cut3] {Remove event if $|m_{bb} - m_h|>$ 20 GeV};

\draw [arrow] (cut1) -- (cut2);
\draw [arrow] (cut2) -- (cut3);
\draw [arrow] (cut3) -- (cut4);
\end{tikzpicture}
\caption{Event selection used to compute the signal-to-background rates.}
\label{fig:selection}
\end{figure}
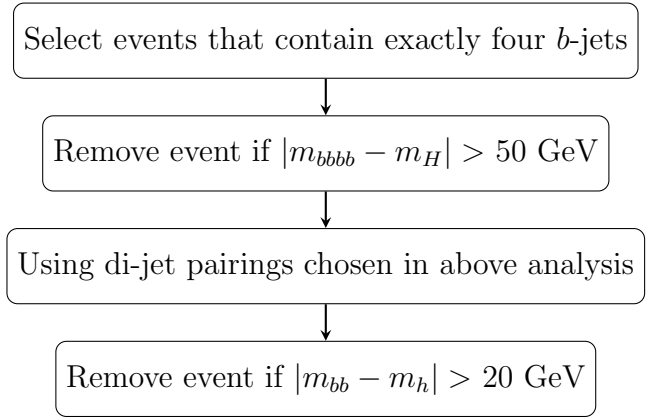

Clearly a good algorithm should not just amplify the signal, but also avoid sculpting the backgrounds. As a final exercise, we perform a calculation of the signal-to-background rates, so as to compare the various jet reconstruction procedures mentioned in this paper also in connection with their performance in dealing with events not coming from our BSM process. In order to do so, we perform the selection
procedure described in Fig.~\ref{fig:selection}. We use the anti-$k_T$ measure throughout but, again, conclusions would not change in case of the CA one. 

\subsubsection{Jet Quality Cuts}
\begin{figure}[h!]
	\begin{center}
	\includegraphics[scale=0.4]{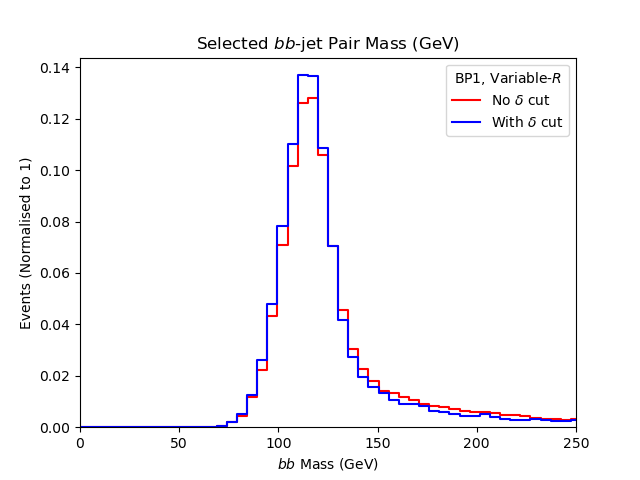}
	\includegraphics[scale=0.4]{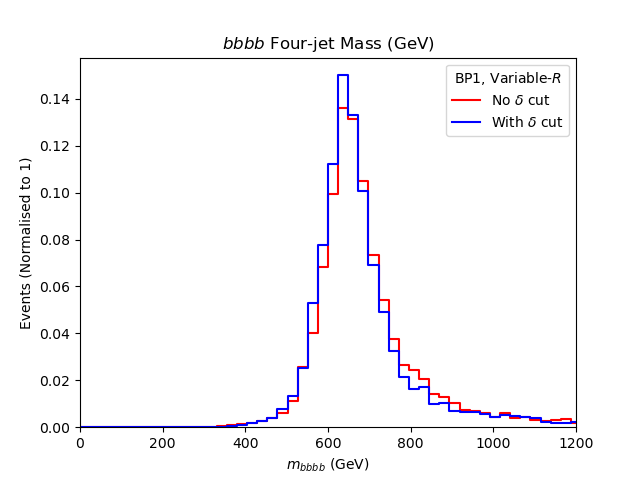}
	 \end{center}
	\caption{Upper panel: The $b$-dijet invariant masses for BP1, with and without the addition of jet quality cuts as defined in Eqs. (\ref{eqn:jet_centre})--(\ref{eqn:delta_cut}). Lower panel: The four $b$-jet invariant mass. Here we have used a value of $\delta=0.05$ for BP1.}
\label{fig:deltacomp_mH700}
\end{figure}

\begin{figure}[h!]
	\begin{center}
	\includegraphics[scale=0.4]{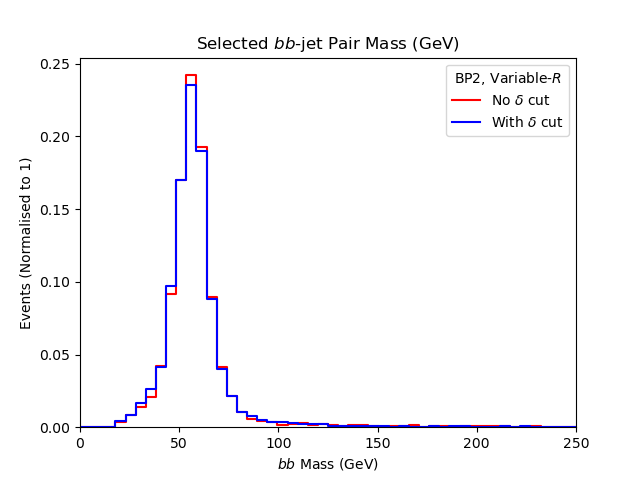}
	\includegraphics[scale=0.4]{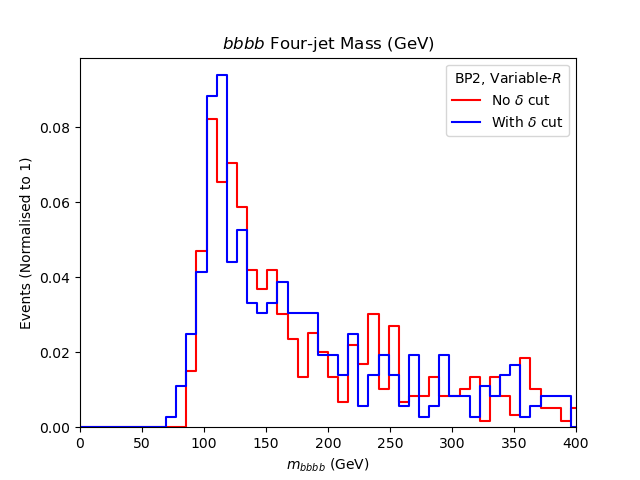}
	 \end{center}
	\caption{Upper panel: The $b$-dijet invariant masses for BP2, with and without the addition of jet quality cuts as defined in Eqs. (\ref{eqn:jet_centre})--(\ref{eqn:delta_cut}). Lower panel: The four $b$-jet invariant mass. Here we have used a value of $\delta=0.1$ for BP2.}
\label{fig:deltacomp_mH60}
\end{figure}

We introduce additional jet quality cuts \cite{Krohn:2009zg} prior to comparing the significance of the signal-to-background characteristics of the two methods. The degree to which the scalar sums of the energy $E_i$ and $p_{Ti}$ constituents of a jet are aligned is taken as a quality measure in \cite{Krohn:2009zg}.  For
\begin{equation}\label{eqn:jet_centre}
{P}_E = \sum_{i\in\mathrm{jet}} E_i \hat{{p}}_i,\ {P}_{p_T} = \sum_{i\in\mathrm{jet}} p_{Ti} \hat{{p}}_i,
\end{equation}
$\hat{{p}}_i$ is the four-momenta of the $i^{th}$ constituent, normalised to unity,  the distance between $P_E$ and $P_{p_T}$ is required to be within a user defined cutoff $\delta$:
\begin{equation}\label{eqn:delta_cut}
\Delta R({P}_E, {P}_{p_T}) < \delta.
\end{equation}
Figs~\ref{fig:deltacomp_mH700}--\ref{fig:deltacomp_mH60} illustrate the utility of jet quality cuts on $b$ dijet and four $b$-jet invariant mass peaks corresponding to $m_h$ and $m_H$. (Here, {we have used a value of $\delta=0.05$ for BP1 and $\delta=0.1$ for BP2.})
Indeed, their application results in higher peaks in both Fig. \ref{fig:deltacomp_mH700} for $m_H=700(125)$ GeV and  Fig. \ref{fig:deltacomp_mH60}  $m_h=125(60)$ GeV.
Finally, we further note that, while there is a hint of signal modification with jet quality cuts (see the lower panel of Fig. \ref{fig:deltacomp_mH700}), the main gains come from the reduction of backgrounds, thereby obtaining higher significances, which we will see in the following section. In this work, we choose the values of the jet quality cut parameter $\delta$ for our BPs following Ref. \cite{Krohn:2009zg}, however, we suggest optimisation of $\delta$ for individual heavy resonance masses to obtain higher significances. Note, alternatively one can also demand the jets to lie in the central region of the detector only or else use the method related to the catchment area of a jet \cite{Cacciari:2008gn} as outlined in reference \cite{Krohn:2009zg}, in order to have better control of the backgrounds.

\subsubsection{Signal Selection}
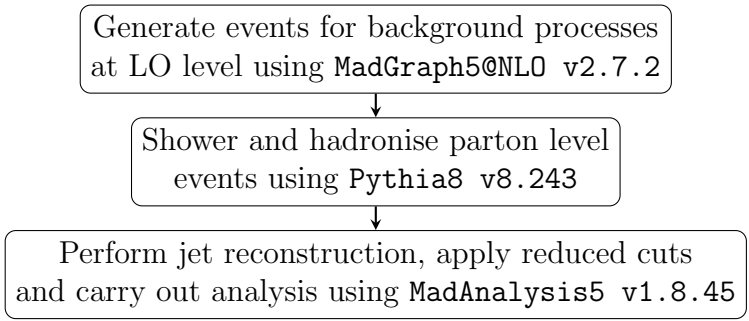
\begin{figure}[!h]
\centering

\begin{tikzpicture}[node distance=1.5cm]
\node (step1) [node,align=center] {Generate events for background processes\\ at LO level using {\tt MadGraph5@NLO v2.7.2}};
\node (step2) [node, align=center,below of=step1] {Shower and hadronise parton level\\ events using {\tt Pythia8 v8.243}};
\node (step3) [node, align=center,below of=step2] {Perform jet reconstruction, apply reduced cuts\\ and carry out analysis using {\tt MadAnalysis5 v1.8.45}};

\draw [arrow] (step1) -- (step2);
\draw [arrow] (step2) -- (step3);
\end{tikzpicture}
\caption{Description of the procedure used to generate and analyse MC events for background processes. }
\label{fig:toolbox_background}
\end{figure}

To carry out this exercise, we generate and analyse $pp \rightarrow b\bar{b}b\bar{b}$, $pp\to Zb\bar b$ and $pp\to t\bar t$ background processes using the toolbox described in Fig.~\ref{fig:toolbox_background} \cite{Alwall:2014hca,Sjostrand:2007gs,Conte:2012fm,Conte:2018vmg}\footnote{We have also checked the non-resonant $pp\to h h\to b \bar b b \bar b$ background for both the BPs. For BP1, in the presence of the described kinematical selections and mass selection criteria described in Tab.~\ref{fig:selection}, the number of events surviving is negligible. For BP2, we get a very small contribution in comparison to the other three backgrounds. Hence, for our MC studies, we do not consider this background.}. Tab.~\ref{tab:signalbackground1} contains the cross sections in pb for signal and the various background processes upon applying the aforementioned cuts and mass selections, including the jet quality cuts.

It is clear from the data obtained that the QCD-induced $pp \rightarrow b\bar{b}b\bar{b}$ process  is the dominant background channel\footnote{In fact, we have computed the full four-jet sample produced by QCD, i.e., including all four-body partonic final states, yet, in presence of the described kinematical selections and $b$-tagging performances, the number of non-$b\bar bb\bar b$ events surviving is negligible \cite{Tannenwald:2020mhq,Behr:2015oqq,Amacker:2020bmn}.}, followed by  $pp\to t\bar t$ and $pp\to Zb\bar b$. Our next step is then to calculate the event rates in order to get the significances for two values of (integrated)  luminosity, e.g.,  ${\mathcal{L}}=$  140 and 300 fb$^{-1}$, corresponding to full Run 2 and 3 data samples, respectively. The event rate ($N$) for the various processes is given by:
\begin{equation}
N \ = \sigma \times \mathcal{L}.
\end{equation}
After the event rates have been calculated, we simply evaluate the significance, $\Sigma$, which is given by (as a function of signal $S$ and respective background $B$ rates)
\begin{equation}
\Sigma = \frac{N(S)}{\sqrt{N(B_{b\bar{b}b\bar{b}})+N(B_{Zb\bar{b}})+N(B_{t\bar{t}})}}.
\end{equation}
Tabs.~\ref{tab:signalbackground4}--\ref{tab:signalbackground5} contains the significances before and after the $K$-factors have been applied. We have used $K = 2$ for the signal \cite{Ravindran:2003um, Harlander:2002wh}, $K =1.5$ for $pp \rightarrow b\bar{b}b\bar{b}$ process \cite{Greiner:2011mp}, $K = 1.4$ for $pp\to t\bar t$ \cite{SM:2010nsa} and $K = 1.4$ for $pp\to Zb\bar b$ \cite{FebresCordero:2009xzo}.
It is then clear from Tabs.~\ref{tab:signalbackground4}--\ref{tab:signalbackground5}
that the variable-$R$ approach provides superior values of significance compared to those obtained from a fixed-$R$, for all choices of $R$ evaluated with and without $K$-factors.   
  
\begin{table*}[htb!]
\centering
\scalebox{1}{
\begin{tabular}{|c|c|c|c|c|}
  \hline
  \multirow{2}{*}{Process} &
    \multicolumn{2}{c}{variable-$R$} \vline &
    \multicolumn{2}{c}{$R=0.4$ } \\ \cline{2-5}
   & BP1 & BP2 & BP1 & BP2 \\
  \hline
  $pp\to H\to hh  \rightarrow b\bar{b}b\bar{b}$ & 2.077 $\times 10^{-4}$& 8.962$\times 10^{-3}$ & 1.254$\times 10^{-4}$ &  3.210$\times 10^{-3}$\\
  \hline
   $pp \rightarrow b\bar{b}b\bar{b}$ & 3.798$\times 10^{-3}$ & 2.131  & 1.651$\times 10^{-3}$ & 9.470$\times 10^{-1}$\\
  \hline
  $pp \rightarrow t\bar{t}$ & 7.973$\times 10^{-4}$ & 2.850$\times 10^{-2}$  & 1.595$\times 10^{-3}$ & 2.217$\times 10^{-2}$ \\
  \hline
  $pp \rightarrow Z b\bar{b}$ & 9.689$\times 10^{-6}$ & 2.627$\times 10^{-2}$ & 3.876$\times 10^{-6}$ & 9.695$\times 10^{-3}$  \\
  \hline
\end{tabular}
}
  \caption{\label{tab:signalbackground1} Cross sections (in pb) of signal and background processes upon enforcing the reduced cuts plus the mass selection criteria $|m_{bbbb}-m_H|< 50$ GeV and $|m_{bb} - m_h|< 20$ GeV for the various jet reconstruction procedures.}
\end{table*}

\begin{table}[!h]
\begin{center}
\scalebox{1}{
\begin{tabular}{ |c|c|c|c| }
 \hline
 & variable-$R$ & $R=0.4$    \\
 \hline
BP1 & 1.145 & 0.823  \\
 \hline
BP2 & 2.268 & 1.214 \\
\hline
\end{tabular}
}
\\
\scalebox{1}{
\begin{tabular}{ |c|c|c|c| }
 \hline
 & variable-$R$  & $R=0.4$     \\
 \hline
BP1 &1.881  & 1.366   \\
 \hline
BP2 & 3.707& 1.984   \\
\hline
\end{tabular}
}
\caption{\label{tab:signalbackground4} Upper panel: Final $\Sigma$ values calculated for signal and backgrounds for ${\cal L}=140$ fb$^{-1}$  upon enforcing the reduced cuts plus the mass selection criteria. Lower panel: Final $\Sigma$ values calculated for signal and backgrounds for ${\cal L}=140$ fb$^{-1}$ with $K$-factors upon enforcing the reduced cuts plus the mass selection criteria.}
\end{center}
\end{table}

\begin{table}[!h]
\begin{center}
\scalebox{1}{
\begin{tabular}{ |c|c|c|c| }
 \hline
 & variable-$R$  & $R=0.4$ \\
 \hline
BP1 & 1.676  & 1.205 \\
 \hline
BP2  & 3.320 & 1.777  \\
\hline
\end{tabular}
}
\\
\scalebox{1}{
\begin{tabular}{ |c|c|c|c| }
 \hline
 & variable-$R$ & $R=0.4$     \\
 \hline
BP1 &2.753  & 2.000   \\
 \hline
BP2 & 5.426& 2.905   \\
\hline
\end{tabular}
}
\caption{\label{tab:signalbackground5} Upper panel: Final $\Sigma$ values calculated for signal and backgrounds for ${\cal L}=300$ fb$^{-1}$  upon enforcing the reduced cuts plus the mass selection criteria. Lower panel: Final $\Sigma$ values calculated for signal and backgrounds for ${\cal L}=300$ fb$^{-1}$ with $K$-factors upon enforcing the reduced cuts plus the mass selection criteria.}
\end{center}
\end{table}

\subsection{Variable-$R$ and Pile-Up}
It has been noted that the nature of variable-$R$, combined with our reduced $p_T$ restrictions, allows for wider cone signal $b$-jets. We therefore perform an analysis of events with Pile-Up (PU) and Multiple Parton Interactions (MPIs), using variable-$R$. As briefly mentioned, in order to perform such a study a proper detector simulation is required. We therefore now employ the use of {\tt Delphes}, passing our hadronised events (simulated in {\tt Pythia8}) through the CMS card (with the same $b$-tagging efficiencies and mistag rates as before). {Specifically, for PU simulations, we have used {\tt Pythia8} to generate soft QCD events. Mixing of these PU events with the signal events is then done within the {\tt Delphes} CMS card, where we have used $\langle N_{PU}\rangle$ = 50 for each hard scattering}. We also perform the same exercise with a fixed cone (anti-$k_T$) of $R=0.4$ to compare.

\begin{figure}[h!]
	\begin{center}
	\includegraphics[scale=0.4]{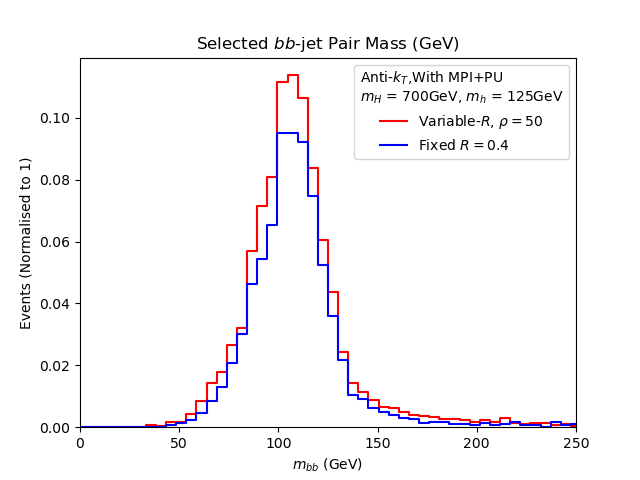}
	\includegraphics[scale=0.4]{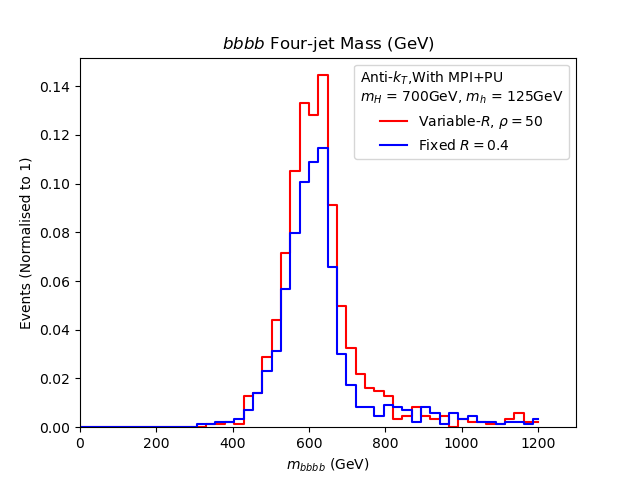}
	 \end{center}
	\caption{Upper panel: The $b$-dijet invariant masses for BP1, using variable-$R$ and fixed-$R$ clustering, when considering the effect of PU and MPIs. Lower panel: The same for the $4b$-jet invariant mass.}
\label{fig:PU_bp1}
\end{figure}

\begin{figure}[h!]
	\begin{center}
	\includegraphics[scale=0.4]{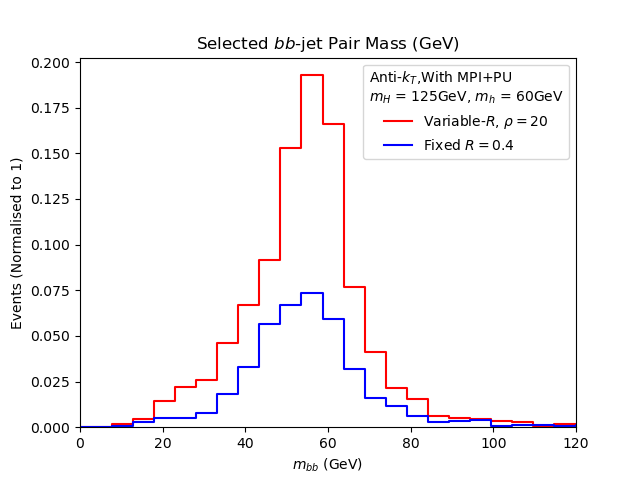}
	\includegraphics[scale=0.4]{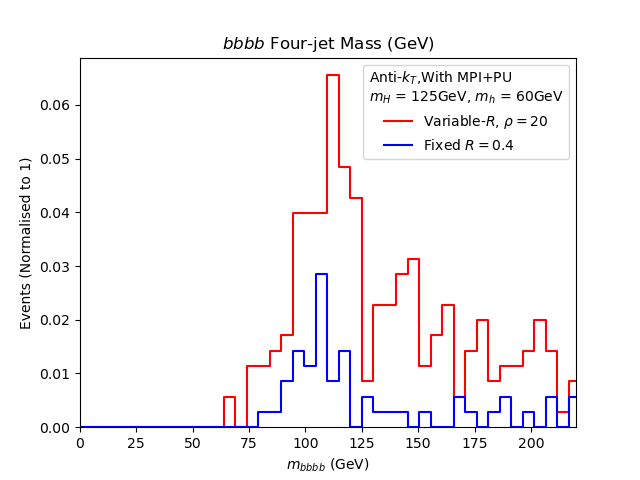}
	 \end{center}
	\caption{Upper panel: The $b$-dijet invariant masses for BP2, using variable-$R$ and fixed-$R$ clustering, when considering the effect of PU and MPIs. Lower panel: The same for the $4b$-jet invariant mass.}
\label{fig:PU_bp2}
\end{figure}

We present in Figs. \ref{fig:PU_bp1}--\ref{fig:PU_bp2} the comparative results for $R=0.4$ and variable-$R$ jet clustering:
herein,  we plot the dijet mass $m_{bb}$ and four-jet mass $m_{bbbb}$ spectra for the signal with PU and MPIs. Note that, with the addition of PU, we have had to use a different value for the variable-$R$ parameter $\rho$, i.e., $\rho=50$. (Having shown the effect of introducing jet quality cuts for background reduction in the previous subsection, we do not use these with PU enhanced events here.)  We see that, with PU added on top of our signal events, many more events are selected following a variable-$R$ jet reconstruction. 

As a final point, we note that a further PU mitigation technique is possible in variable-$R$, which is in the values chosen for the $R_{\text{min/max}}$ variables. Clearly if, for some particular process, one discovers that using a variable-$R$ sweeps in too much extra `junk' into the jets, a simple reduction of $R_{\text{max}}$ is always possible. (Notice that, in order to reduce the 
contamination due to PU and MPIs,  one can always use grooming techniques such as filtering \cite{Butterworth:2008iy}, trimming \cite{Krohn:2009th}, pruning \cite{Ellis:2009su}, mass-drop \cite{Butterworth:2008iy}, modified mass-drop \cite{Dasgupta:2013ihk} and soft drop \cite{Larkoski:2014wba}, however, this is beyond the scope of this paper.) 

\subsection{Other Variable-$R$ Studies}
Before concluding, we review here some other studies from the literature that use a variable-$R$ reconstruction algorithm.

We note that, while the leading $b$-jet has an $R_{\text{eff}}$ roughly in line with expected values ($R_{\text{eff}} \simeq 0.5$), the lowest $p_T$ $b$-jets have large cone sizes ($R_{\text{eff}} > 1.0$), risking potential contamination from additional radiation. This effect is discussed in \cite{Lapsien:2016zor}. We do not implement any vetoes to remedy this effect. Despite this, our results suggest that the variable-$R$ clustering algorithm shows an improvement over standard anti-$k_T$. 

There have been other studies utilising variable-$R$ methods for physics searches, such as in the highly boosted object tagging of $hh \rightarrow bb\bar{b}\bar{b}$ decays in \cite{ATLAS:2016hcf}.  The variable-$R$ algorithm was also deployed  in \cite{Lapsien:2016zor} to analyse heavy particle decays. In both examples, an improvement over current fixed-$R$ methods is present when using variable-$R$, which is in line with our findings.

As a final word on using variable-$R$ jet reconstruction in experiments, we discuss its use in relation to $b$-tagging performance. In particular, the studies of Refs.~\cite{ATLAS:2017juw, ATLAS:2020udg} explore the possibility of Higgs to $b$-jet tagging at ATLAS using variable-$R$ techniques. Specifically, since these studies deal with boosted topologies, focusing on fat $b$-jet substructure, the advantage of applying these techniques in a non-boosted regime is yet to be determined. 

\section{Conclusions}
In this paper, we have assessed the potential scope of the LHC
experiments in accessing BSM Higgs signals induced by cascade decays
of the 125 GeV SM-like Higgs state discovered in July 2012 into two lighter Higgs states or indeed of a heavier one into pairs of it.
{The prototypical production and decay channel that we have used is  
    $gg,q\bar q$ $\to$ $H\to hh$. 
Two mass benchmarks are considered; in the first, $h$ is the SM-like Higgs state
and $H$ is a heavier BSM Higgs state, with mass greater than $2m_h$, so as to induce resonant
production and decay, thereby enhancing the overall rate.
Conversely, the second mass benchmark takes $H$ to be the SM-like Higgs, with $h$ as a lighter Higgs with mass less than $m_H/2$,  leading to the same resonance effects.}
Either light Higgs boson would decay to $b\bar b$ pairs, eventually leading to a four $b$-jet 
signature, largely independently of the BSM construct hosting it.
 
The four-jet signature is extremely difficult to detect at the LHC,
owing to  substantial hadronic background. Thus,
$b$-tagging techniques are to be exploited in order to make such a
signal visible. These taggers are most efficient when $b$-jets have a large transverse momentum, say, at least 20 GeV.  At this scale there is a significant loss of signal events if the BSM Higgs mass is in the sub-60 GeV range. The conflict between tagging efficiency and signal retention poses a problem.  Hence,
if one intends to maximise sensitivity to this hallmark signature of
BSM physics, a thorough reassessment of the current Run 2 approaches
is mandated for and especially so in view of the upcoming Run 3. 

The first message we deliver is that, with current $p_T$ cuts on final state $b$-jets, using a fixed-$R$ jet reconstruction and tagging procedure will lead to a poor signal visibility, with a majority of signal $b$-jets being lost. We instead presented a reduced cutflow, based on existing $b\bar b\mu^+\mu^-$ analyses, and showed that this indeed provides a window onto $gg \rightarrow H \rightarrow hh \rightarrow bb\bar{b}\bar{b}$ signals with $m_H = 125$ GeV and $m_h < \frac{m_H}{2}$. 

Additionally, and perhaps more remarkably, we also test\-ed a variable-$R$ reconstruction approach on events with this reduced cutflow and showed a significant improvement in signal yield as well as signal-to-background rates. We notice that in final states of this kind, the signal $b$-jets have a wide range of $p_T$ and hence varied spread of constituents. Using a fixed cone of a standard size ($R=0.4$) constructs well higher $p_T$ jets in an event but does not capture much of the wider angle radiation from lower $p_T$ jets. This leads to two issues. Firstly, it will prove difficult to accurately construct $m_h$ and $m_H$ in the two- and four-jet invariant masses. Secondly, these jets will more often be lost due to kinematic cuts. 

We have obtained all of the above in presence of a sophisticated MC event simulation, based on exact scattering MEs, state-of-the-art parton shower, hadronisation and $B$-hadron decays as well as a mainstream detector simulation, also including PU and MPIs. Given the results of our analysis, we recommend a more thorough detector level analysis which is being undertaken for a variety of different high $b$-jet multiplicity scenarios.  This analysis should explore whether, on the one hand, a shift to variable-$R$ jet clustering could be implemented and, on the other, would improve upon current signal significance limitations using fixed-$R$ jet reconstruction. In fact, while we have quantitatively based our case on the example of the 2HDM-II, our procedure can identically be used in other BSM constructs featuring the same Higgs cascade decays.

\section*{Acknowledgments}
SM is supported also in part through the NExT Institute and the STFC Consolidated Grant No. ST/L000296/1. BF is funded by the DISCnet SEPnet scholarship scheme. The work of AC is funded by the Department of Science and Technology, Government of India, under Grant No. IFA18-PH 224 (INSPIRE Faculty Award).
We all thank G.P. Salam for useful advice. We also give thanks to Benjamin Fuks, Eric Conte and others in the MadAnalysis5 team for their assistance with technical queries.
BF and SJ acknowledge the use of the IRIDIS High Performance Computing Facility, and associated support services at the University of Southampton, in the completion of this work.



\clearpage
\thispagestyle{empty}

\end{document}